\documentclass[fleqn,twoside]{article}
\usepackage{espcrc2}
\usepackage{epsfig}
\usepackage{amssymb}

\newcommand{\be}{\begin{equation}}
\newcommand{\ee}{\end{equation}}
\newcommand{\bea}{\begin{eqnarray}}
\newcommand{\eea}{\end{eqnarray}}

\newcommand{\squeeze}{\!\!\!\!\!\!\!\!}
\newcommand{\squ}{\!\!\!\!}
\begin{document}

\title{Functional Forms for Lattice Correlators at Small Times}
\author{Danielle Blythe\address {Department of Physics, University of Wales Swansea, U.K.}}
\begin{abstract}The analytic form of the lattice quark propagator is
used to derive the functional form for short distance mesonic
correlators. These are then used to calculate ``Continuum Model''
Ansatz\"e which comprise of a pole, representing the ground state, plus
a contribution for the excited states, coming from the short distance
behaviour. These are compared to Monte Carlo data.
\end{abstract}
\maketitle 

\section{Introduction}
On the lattice, the usual method for studying
hadronic physics is to determine the two-point correlation functions of
hadronic operators.
\begin{eqnarray}\label{eqn:hadcorrfn}
G_2 ^{\Gamma} (t)& =
&\sum_{\underline{x}} \langle 0| T
\{J_{\Gamma}(\underline{x},t)\overline{J}_{\Gamma}(\underline{0},0)\}|0\rangle \nonumber\\
&\longrightarrow& e^{-Mt} \quad\mathrm{as}\quad t\rightarrow\infty,
\end{eqnarray}
where $M$ is the mass of the ground state, and $\Gamma$ is channel dependent.
The presence of the decaying exponential has led to investigations
being focused on the large Euclidean time
``tails'' of the two-point functions, and data at small times being discarded.

The question was asked, ``Is there a way of obtaining information from
short times?''. The answer was ``Yes'' -  a QCD Sum Rules approach.
Lattice QCD allows the determination of correlation functions deep
within the non-perturbative region. QCD Sum Rules (QCDSR) can be
utilised in the near-perturbative regime where the excited states
cannot be ignored. Both are techniques which have been widely employed
to deepen the understanding of strongly interacting systems, and have
been combined in the work of Leinweber\cite{lw1,lw2}, and Allton and
Capitani \cite{long}. 

In this study, we extend the work of \cite{lw1,lw2,long} by applying
the analysis with a continuum limit expansion of the \textit{lattice}
quark propagator, derived in \cite{paladini}.

The lattice data used is UKQCD generated, and is (quenched)
Wilson at $\beta=6.0$ with a volume of $16^3 \times 48$.
\vspace{1cm}
\section{QCDSR Continuum Model}
In this section we introduce the QCDSR-CM and apply it to the continuum
quark propagator. The QCDSR-CM replaces a discrete spectrum of excited
states with a
continuous spectrum of spectral density $\rho$, and
introduces a sharp threshold in the energy scale, $s_0$, marking the onset of
the continuum.\\
The basic object studied is the continuum OPE for QCD quark propagator:
\begin{equation}
S^{OPE}_q(x) =
           \frac{\gamma \cdot x}{2 \pi^2 x^4} +
           \frac{m}{(2 \pi x)^2}            -
           \frac{\langle : \!\overline{q}q\! : \rangle}{2^2 3}\!
           + \cdots,
\label{eqn:qprop}
\end{equation}
where $m$ is the quark mass.
This is substituted into the (Wick-contacted) 2-point function at zero-momentum:
\begin{equation}
G_2(t)^{OPE}\!\!=\!\! \int\!\! d^3\!x\:\:\:\:Tr \!\!\left\{S_q(x)^{OPE}\Gamma
S_q(-x)^{OPE}\Gamma\right\}.
\label{eqn:g2}
\end{equation}
It is useful to express the time-sliced correlation function,
$G_2(t)^{OPE}$, in the spectral representation:
\begin{equation}
G_2(t)^{OPE}= \int_0^{\infty} \rho(s)^{OPE}\:
e^{-st}\:\mathrm{ds},
\end{equation}
and the OPE spectral density is
calculated by means of an inverse Laplace transform of $G_2^{OPE}$. We
invoke the QCDSR-CM  by setting the threshold $s_0$ in the energy
scale, so that the excited states' contributions to $G_2(t)$ is given
by only the energies above that scale,
\begin{equation}\label{eqn:deltafunction}
\rho(s) = Z\:\delta(s-M) + \xi
\:\theta(s-s_0)\:\rho(s)^{OPE}.\end{equation}
We have to include the ground state in order to obtain the full correlation
function. This is included as a $\delta$ function in $\rho(s)$ in
Eq.(\ref{eqn:deltafunction}). So:
\begin{eqnarray}
  G_2(t)=Ze^{-Mt}+\xi\int_{s_0}^\infty
\rho(s)^{OPE}\:e^{-st}\:\mathrm{ds}.
\end{eqnarray}
There are four parameters in the fitting ansatz:\\
Z $\longrightarrow$ normalisation of ground state\\
M $\longrightarrow$ ground state mass\\
$\xi$ $\longrightarrow$ normalisation of excited states, introduced to allow
for lattice distortions \cite{lw1}\\
$s_o$ $\longrightarrow$ continuum threshold.\\
The OPE expansions for the degenerate mesons and baryons using the
continuum quark propagator are listed in \cite{lw1,long}. The analysis
was extended to non-degenerate mesons in \cite{mejonchris}. 
\section{Wilson Quark Propagator}
In terms of loop-divergent behaviours, a coordinate space formulation of a
lattice theory, has the benefit over
its momentum analogue of highlighting the short-distance phenomena which are
the origin of the poles. In the
continuum, the analytic expression for a fermion propagator in position
space is well known, Eq.(\ref{eqn:qprop}). On the lattice
however, the standard representation involves integrals over Bessel
functions, and proved difficult to analyse in
the continuum limit. L\"{u}scher and Weisz
successfully analysed the massless propagator in
$4$-dimensions \cite{luscherweisz}.
While in \cite{bcp} the mass dependence of the propagator at
$x=0$ was derived. However, in
\cite{paladini}, the asymptotic expansion for the modified Bessel function
of the first kind, $I_n$, which appears in the expression for the
lattice propagator, (using Schwinger's integral representation), 
\begin{eqnarray}
S_q(x)^{WF}\!\!\!\squ&=&\squeeze
\sum_{i=even}\!\!\left\{\!m\!-\!ia^{-1}\!\!\sum_{\mu=1}^4 \!\gamma_\mu \sin\left[a(-i\partial_\mu \!+k_\mu^i)\right]\right. \nonumber\\
\squ&+&\squ\left. ra^{-1} \sum_{\mu=1}^4 \left\{1-\cos\left[a(-i\partial_{\mu} + k_\mu^i)\right] \right\}\right\}\nonumber\\
\squ&\times&\squ \int_0^{\infty} d\alpha \:e^{-m^2\alpha} \exp\left\{ -\alpha[4mra^{-1}]\right.\nonumber\\
\squ&\times&\squ\left.\sum _{\mu=1}^4 \sin^4 \left(\frac{-ia\partial_{\mu}}{2}\right)\right\}\nonumber\\
\squ&\times&\squ \exp \left\{-\alpha \left(\frac{r^2a^{-2}}{4}\right)\left[ \sum_{\mu=1}^4
\sin^2 (-ia\partial_{\mu})\right.\right.\nonumber\\
\squ&+&\squ\left.\left. 4 \sum_{\mu=1}^4 \sin^4 \left(\frac{-ia\partial_{\mu}}{2}\right)\right]^2 \right\} \nonumber\\
\squ&\times&\squ\prod_{\mu=1}^4 \left(\frac{1}{2a} \exp \left\{-\frac{\alpha}{2} [a^{-2} + mra^{-1}]\right\}\right.\nonumber\\
&\times&\squ\left. I_{\frac{x_\mu}{2a}}\left\{\frac{\alpha}{2}[a^{-2}+mra^{-1}]\right\}\right), \!\!\!\!\!\!\!
\eea
 was calculated, leading to the following expression for the Wilson lattice
propagator:
\begin{eqnarray} \label{eqn:expandedlattprop}
(4\pi)^2S_q(x)^{WF}\!\!\!\!\!\!\!&=&\!\!\!\!\!
\frac{4m^2}{(x^2)^\frac{1}{2}} \left( 1 +
\frac{ram}{2} \right)\!\!K_1\!\!\left[
m(x^2)^{\frac{1}{2}}\right]\nonumber\\\!\!\!\!\!\!\!\!\!&+&\!\!\!\!\! 2ram^4 K_2\left[ m{(x^2)}^\frac{1}{2}\right]
\nonumber\\
\!\!\!\!\!\!\!\!&+&\!\!\!\!2ram^4 \frac{\gamma\cdot x}{(x^2)^\frac{1}{2}} K_1\left[
m{(x^2)}^\frac{1}{2}\right]\nonumber\\\!\!\!\!\!\!\!\!\!&+&\!\!\!\!
4m^2(1 + ram) \frac{\gamma\cdot x}{x^2}K_2\!\!\left[
m{(x^2)}^\frac{1}{2}\right]\nonumber\\\!\!\!\!\!\!\!\!\!&+&\!\!\!\! 
\mathcal{O}(\mathit{a^2}),
\end{eqnarray}
where $K_1$ and $K_2$ are modified Bessel functions of the second kind.
This expression was then used in a QCDSR-CM style approach following
the procedures outlined in section 2.
\begin{figure}[htp]
\begin{center}
\epsfig{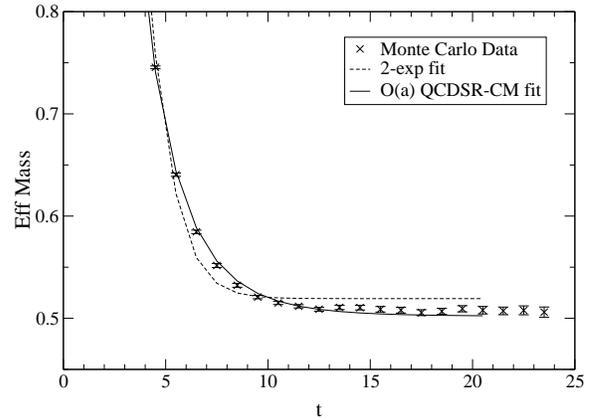}
\end{center}
\vspace{-10mm}
\label{fig}
\caption{Effective mass plot for the 2-exponential and
${\cal O}(a)$ QCDSR-CM fits
for the vector channel at $\kappa = 0.1530$.}
\vspace{-3mm}
\end{figure}
%
 \begin{table}[*htbp]
 \begin{center}
 \begin{tabular}{llll}
\hline
Fit & \multicolumn{1}{c}{Z} & \multicolumn{1}{c}{M} & $\chi^2/\rm{d.o.f}$ \\
Type &&&\\
\hline
\multicolumn{4}{c}{\bf Pseudoscalar} \\
\multicolumn{4}{c}{ $\kappa = 0.1550$} \\
%
%
1-exp  & 1.81(6)$\times 10^{-2}$ & 0.2986(15) & - \\
2-exp  & 1.91(4)$\times 10^{-2}$ & 0.3014(12) & 10/15 \\
${\cal O}(a)$  & 1.7(3)$\times 10^{-2}$ & 0.294(3) & 8/15 \\
\hline
\multicolumn{4}{c}{ $\kappa = 0.1540$} \\
%
%
1-exp  & 2.12(5)$\times 10^{-2}$ & 0.3645(11) & - \\
2-exp  & 2.28(4)$\times 10^{-2}$ & 0.3685(10) & 24/15 \\
${\cal O}(a)$  & 1.6(14)$\times 10^{-2}$ & 0.356(14) & 18/15 \\
\hline
\multicolumn{4}{c}{ $\kappa = 0.1530$} \\
%
%
1-exp  & 2.46(5)$\times 10^{-2}$ & 0.4236(9) & - \\
2-exp  & 2.67(3)$\times 10^{-2}$ & 0.4281(8) & 42/15 \\
${\cal O}(a)$  & 2.1(4)$\times 10^{-2}$ & 0.416(3) & 30/15 \\
\hline
\hline
\multicolumn{4}{c}{\bf Vector} \\
\multicolumn{4}{c}{ $\kappa = 0.1550$} \\
%
%
1-exp  & 6.7(5)$\times 10^{-3}$ & 0.426(4) & - \\
2-exp  & 8.5(2)$\times 10^{-3}$ & 0.440(2) & 160/15 \\
${\cal O}(a)$  & 5.9(2)$\times 10^{-3}$ & 0.419(3) & 30/15 \\
\hline
\multicolumn{4}{c}{ $\kappa = 0.1540$} \\
%
%
1-exp  & 8.6(3)$\times 10^{-3}$ & 0.467(3) & - \\
2-exp  & 1.03(2)$\times 10^{-2}$ & 0.479(2) & 210/15 \\
${\cal O}(a)$  & 7.5(2)$\times 10^{-3}$ & 0.461(2) & 37/15 \\
\hline
\multicolumn{4}{c}{ $\kappa = 0.1530$} \\
%
%
1-exp  & 1.05(3)$\times 10^{-2}$ & 0.509(2) & - \\
2-exp  & 1.25(2)$\times 10^{-2}$ & 0.5193(13) & 240/15 \\
${\cal O}(a)$  & 9.3(2)$\times 10^{-3}$ & 0.5020(15) & 41/15 \\
 \hline
 \end{tabular}
 \end{center}
 \caption{Values of $Z$, $M$ and $\chi^2$ for the fitting ansatz\"e.}
 \label{table}
 \end{table}

\vspace{-10mm}
\section{Wilson Correlators}
In this section, the work of \cite{lw1,lw2,long} and \cite{paladini} is
combined by applying a QCDSR-CM approach
to the continuum limit expansion of the Wilson lattice propagator,
i.e. we use $S_q^{OPE} \equiv S_q^{WF}$ in Eq.(\ref{eqn:g2}).

To obtain a closed expression for $G_2$ we
first expand the modified Bessel functions for short
distances and then perform the spatial integral.
This procedure was
performed for all mesonic channels, with general (non-degenerate)
constituent quarks, and for the momenta, $\underline{p}=\underline{0}$ and
$\underline{p}=(1, 1, 1)$. We obtain, e.g., for the zero-momentum
pseudoscalar channel the following expression:
\bea
G_2^{PS}(t)^{OPE}&=& -\frac{3}{4\pi^2t^3}(1+2ram)
\nonumber\\&+&
\frac{3m^2}{4\pi^2t}(1-ram).\eea

Values for the $Z$ and $M$ fitting parameters together with the $\chi^2$
for the pseudoscalar and vector channels are given in Table~\ref{table}.
The QCDSR-CM fits for the Wilson correlators are labelled by ${\cal O}(a)$.
The fitting window for the {\em 1-exp} fit was $t=[12,20]$ and for the
{\em 2-exp} and ${\cal O}(a)$ fits was $t=[2,20]$. Note that the 
${\cal O}(a)$ fits perform better than the {\em 2-exp} fit in every
case, and that the improvement is more significant for the vector
channel particularly as the mass increases.

\section{Conclusions}
In this work we have derived lattice mesonic correlation functions for the
Wilson action, including
$\mathcal{O}(\mathit{a})$ effects. Our Ansatz\"{e} fit lattice Monte Carlo data better than
conventional ``2-exponential'' fits.
Future work includes fitting of heavy mesons, and deriving analogous
formulae for the Clover action.
\section{Acknowledgements}
The author would like to thank Craig McNeile and Chris Allton.

\end{document}